\documentclass[sigconf]{acmart}

\renewcommand\footnotetextcopyrightpermission[1]{} 

\usepackage{amsmath,amssymb,amsfonts}
\usepackage{hyperref}
\hypersetup{colorlinks=true}
\def\equationautorefname~#1\null{%
  Equation~(#1)\null
}
\usepackage{algorithm}
\usepackage[noend]{algpseudocode}
\usepackage{colortbl}
\usepackage{xcolor}
\usepackage{colortbl}%

\usepackage{capt-of,lipsum}
\definecolor{myyellow}{HTML}{FFD966} 
\definecolor{mypink}{HTML}{FF99CC} 
\definecolor{mycolor}{rgb}{0.95, 0.985, 0.93}
\doublerulesepcolor{mycolor}
\usepackage{xspace}
\usepackage{array}
\usepackage{color}
\usepackage{enumitem}

\setlist[itemize]{%
  leftmargin=15pt,
  topsep=3pt,
  itemsep=1pt,
  parsep=0pt,
  partopsep=0pt
}
\setlist[enumerate]{%
  leftmargin=15pt,
  topsep=3pt,
  itemsep=1pt,
  parsep=0pt,
  partopsep=0pt
}

\usepackage{pifont}
\usepackage{xcolor}
\usepackage{listings}
\usepackage{makecell}
\usepackage{multirow}
\usepackage{subcaption}
\usepackage{booktabs}
\usepackage[most]{tcolorbox}
\usepackage{comment}
\usepackage{tikz}
\usepackage{diagbox}
\usepackage[normalem]{ulem}

\definecolor{keywordcolor}{rgb}{0.13,0.29,0.53}
\definecolor{stringcolor}{rgb}{0.31,0.60,0.02}
\definecolor{commentcolor}{rgb}{0.56,0.35,0.01}
\definecolor{backcolour}{rgb}{0.95,0.95,0.92}
\AtEndPreamble{
	\usepackage{hyperref}

        \def\equationautorefname{Equation}
        
	\hypersetup{
		colorlinks = true,
		linkcolor = purple,
		anchorcolor = purple,
		citecolor = purple,
		filecolor = purple,
		urlcolor = purple
	}
}

\newcommand{\toolname}{\textsc{MATP}\xspace}
\newcommand{\proofwriter}{\textsc{ProofWriter}\xspace}
\newcommand{\prontoqa}{\textsc{PrOntoQA-OOD}\xspace}
\newcommand{\folio}{\textsc{FOLIO}\xspace}
\newcommand{\vampire}{\textsc{Vampire}\xspace}

\definecolor{mGray}{rgb}{0.5,0.5,0.5}
\newcommand{\commentstyle}[1]{\textcolor{mGray}{#1}}

\definecolor{darkgreen}{RGB}{100,140,100}
\definecolor{newgreen}{RGB}{52,168,83}
\AtBeginDocument{%
  \providecommand\BibTeX{{%
    \normalfont B\kern-0.5em{\scshape i\kern-0.25em b}\kern-0.8em\TeX}}}

\copyrightyear{2026}
\acmYear{2026}
\setcopyright{acmlicensed}
\acmConference[ICSE '26]{48th IEEE/ACM International Conference on Software Engineering}{April 12-18, 2026}{Rio de Janeiro, Brazil}
\acmBooktitle{48th IEEE/ACM International Conference on Software Engineering (ICSE '26), April 12-18, 2026, Rio de Janeiro, Brazil}
\begin{document}
\begin{sloppypar}

\title{Beyond Correctness: Exposing LLM-generated Logical Flaws in  Reasoning via Multi-step Automated Theorem Proving}


\author[X Zheng]{Xinyi Zheng}
\authornote{Both authors contributed equally to this research.}
\affiliation{%
  \institution{\emph{Huazhong University of Science and Technology}}
  \city{Wuhan}           
  \country{China}
}
\email{xinyi_zheng@hust.edu.cn}

\author[N Li]{Ningke Li}
\authornotemark[1]
\affiliation{%
  \institution{\emph{National University of Singapore}}          
  \country{Singapore}
}
\email{ningke_li@u.nus.edu}

\author[X Luan]{Xiaokun Luan}
\affiliation{%
  \institution{\emph{Peking University}}
  \city{Beijing}           
  \country{China}
}
\email{luanxiaokun@pku.edu.cn}

\author[K Wang]{Kailong Wang}
\authornote{Kailong Wang~(wangkl@hust.edu.cn) is the corresponding author.}
\affiliation{%
  \institution{\emph{Huazhong University of Science and Technology}}
  \city{Wuhan}           
  \country{China}
}
\email{wangkl@hust.edu.cn}

\author[L Shi]{Ling Shi}
\affiliation{%
  \institution{\emph{Nanyang Technological University}}        
  \country{Singapore}
}
\email{ling.shi@ntu.edu.sg}

\author[M Sun]{Meng Sun}
\affiliation{%
  \institution{\emph{Peking University}}
  \city{Beijing}           
  \country{China}
}
\email{sunm@pku.edu.cn}

\author[H Wang]{Haoyu Wang}
\affiliation{%
  \institution{\emph{Huazhong University of Science and Technology}}
  \city{Wuhan}           
  \country{China}
}
\email{haoyuwang@hust.edu.cn}



\begin{abstract}
Large Language Models (LLMs) have demonstrated impressive reasoning capabilities, leading to their adoption in high-stakes domains such as healthcare, law, and scientific research. However, their reasoning often contains subtle logical errors masked by fluent language, posing significant risks for critical applications. 
While existing approaches like fact-checking, self-consistency methods, and rule-based validation provide partial solutions, they fail to detect complex logical flaws in multi-step reasoning. 

To overcome these challenges, we present \toolname, an evaluation framework for systematically verifying LLM reasoning via Multi-step Automatic Theorem Proving. \toolname translates natural language reasoning into First-Order Logic (FOL) and applies automated theorem provers to assess step-by-step logical validity. 
This approach identifies hidden logical errors and provides fine-grained classifications of reasoning correctness. 
Evaluations on a benchmark comprising 10,830 reasoning instances generated by 10 LLMs across tasks from \prontoqa, \proofwriter, and \folio show that \toolname surpasses prompting-based baselines by over 42 percentage points in reasoning step verification.
It further reveals model-level disparities, with reasoning models generating more logically coherent outputs than general models. These results demonstrate \toolname's potential to enhance the trustworthiness of LLM-generated reasoning.

\end{abstract}





\maketitle
\section{Introduction}

Large Language Models (LLMs) have achieved remarkable progress in natural language understanding and generation, demonstrating strong capabilities in reasoning~\cite{qiao-etal-2023-reasoning, huang-chang-2023-towards, ahn-etal-2024-large}, inference~\cite{wan2024efficient}, and decision-making~\cite{Yang2023FoundationMF, Eigner2024DeterminantsOL}. Their success has driven their adoption in high-stakes domains including healthcare~\cite{Clusmann2023}, law~\cite{LAI2024181}, finance~\cite{10.1145/3604237.3626869}, and scientific research~\cite{zhang-etal-2024-comprehensive-survey, 10.1145/3581784.3613215}, where reliable reasoning is essential. 
However, safe deployment remains challenging: despite producing fluent and seemingly coherent outputs, LLMs often exhibit subtle logical errors, contradictions, or unsupported inferences~\cite{li-etal-2024-reason, payandeh-etal-2024-susceptible, mndler2024selfcontradictory} that could lead to serious consequences like misdiagnoses, flawed legal judgments, or erroneous research conclusions. 
Recent efforts to address these limitations include Large Reasoning Models (LRMs) that augment LLMs with neuro-symbolic AI, reinforcement learning, and causal inference. Yet Apple's recent analysis~\cite{illusion-of-thinking} reveals that even LRMs struggle with complex tasks, often relying on shallow heuristics rather than robust reasoning. Moreover, LLM reasoning remains vulnerable to adversarial attacks, where minor input perturbations can systematically corrupt reasoning chains~\cite{gan-etal-2024-reasoning, wu-etal-2024-decot}.

Existing research has largely focused on detecting flawed reasoning in LLM outputs, but existing methods face key limitations. Fact-checking approaches~\cite{tam-etal-2023-evaluating, chen2023felm, Augenstein2024} verify factual accuracy using external knowledge bases but often miss logical errors, especially when reasoning involves novel or implicit fact combinations. Self-consistency methods~\cite{wang2023selfconsistency} reduce random mistakes via majority voting across outputs, yet fail to catch systematic errors repeated across generations. Rule-based and post-hoc validation~\cite{Xie2025LogicRLUL, krishna2023post, servantez-etal-2024-chain} can flag simple contradictions but struggle with complex, multi-step logic, offering only limited coverage of reasoning flaws.

Prior works~\cite{NEURIPS2023_8e9c7d4a,olausson-etal-2023-linc,pan2023logiclm}, using formal methods to enhance LLM reasoning, provide valuable insights.
We consider applying formal methods to evaluate, rather than enhance, LLM reasoning quality as a promising direction for rigorous analysis.
Despite their theoretical appeal, several intertwined challenges remain. 
\textbf{Challenge\#1: absence of explicit logical structure in LLM-generated reasoning.} 
Unlike formal reasoning systems, LLMs generate reasoning via probabilistic pattern matching rather than structured inference, resulting in verbose and implicit reasoning chains.
Formalizing such unstructured reasoning into precise logical representations is highly challenging, especially compared with Logic-LM~\cite{pan2023logiclm}, which only formalizes predefined reasoning tasks.
\textbf{Challenge\#2: subtle and silent reasoning errors within LLMs.}  
A single incorrect or unsupported inference step can silently contaminate subsequent reasoning, producing conclusions that appear plausible yet are fundamentally unsound. 
Identifying and isolating such errors is difficult because it requires precise tracking and validation of logical dependencies throughout an entire reasoning chain.

\begin{figure}[t]
    \centering\vspace{-0.3cm}
    \includegraphics[width=\linewidth]{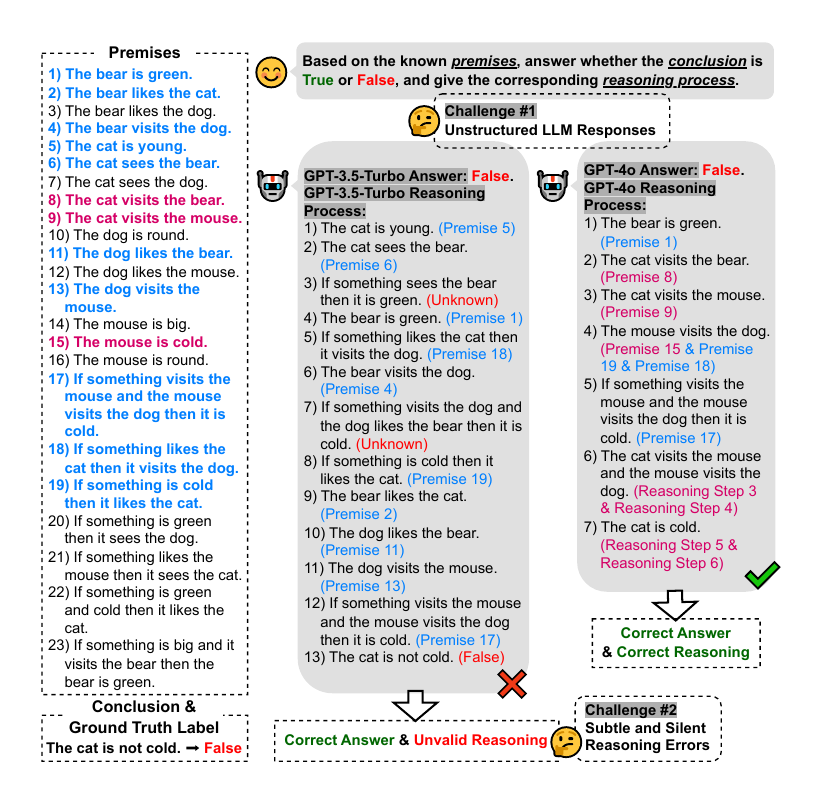}\vspace{-0.5cm}
    \caption{Motivating Example. Premises used by GPT-3.5-Turbo are marked in \textcolor[HTML]{007FFF}{Bright Blue}; GPT-4o includes these and adds \textcolor[HTML]{CC0066}{Deep Magenta} premises. GPT-3.5-Turbo’s reasoning is invalid, while GPT-4o, leveraging Premises 8, 9, and 15, yields the correct answer with a valid reasoning process.}\vspace{-0.3cm}
    \label{fig:motivation}
\end{figure}

To address the challenges outlined above, we propose Multi-step Automatic Theorem Proving~(\toolname), a post-hoc evaluation framework that enables fine-grained analysis of LLM-generated reasoning.
\toolname leverages the natural language understanding capabilities of LLMs to convert both reasoning tasks and model-generated reasoning chains into unified First-Order Logic~(FOL) representations in a single pass~(\textbf{Challenge \#1}), enabling structured verification via Automated Theorem Prover~(ATP). 
It subsequently verifies the correctness of each reasoning step and searches for a valid proof path within the reasoning chain that logically entails the final conclusion, thereby distinguishing genuine deductive reasoning from mere information regurgitation~(\textbf{Challenge \#2}). 
Based on the verification outcomes, \toolname performs fine-grained classification of reasoning chains—identifying whether a chain is fully valid, partially flawed, or logically inconsistent.
This classification jointly considers answer correctness, step-level validity, and proof-path soundness, moving beyond binary accuracy metrics to provide deeper insights into reasoning quality.
Through this design, \toolname provides a more interpretable and reliable assessment of LLM reasoning performance in complex deductive tasks.

\noindent\textbf{Results and Findings.}
We evaluate \toolname{} through extensive experiments on two datasets: (1) a benchmark comprising reasoning tasks from \prontoqa{}~\cite{PrOntoQAOOD}, \proofwriter{}~\cite{tafjord2021proofwriter}, and \folio{}~\cite{han2024folio}, along with LLM-generated responses from 10 mainstream LLMs, including five general and five reasoning models, totaling 10,830 instances; and (2) a classification dataset based on \prontoqa{}, constructed by applying controlled perturbations to standard reasoning chains.
Across the benchmark dataset, \toolname{} converts natural language sentences to FOL representations with over 99\% accuracy on syntactically simpler tasks from \prontoqa{} and \proofwriter{}, while maintaining relatively strong performance on the more linguistically complex tasks from \folio{}. 
In verifying individual reasoning steps, \toolname{} consistently outperforms prompting-based baselines by more than 42 percentage points in macro F1-score.
On the classification dataset, \toolname{} achieves 99\% classification accuracy, substantially outperforming all baselines and demonstrating its ability to distinguish fine-grained reasoning types, though occasional misclassifications remain.

Based on \toolname{}’s evaluation and classification of all LLM-genera\-ted responses in the benchmark dataset, we observe that reasoning models—especially DeepSeek-R1—produce more logically coherent reasoning chains than general models, despite similar answer accuracy. 
Moreover, most evaluation failures originate from errors in the NL2FOL translation stage. We further identify and summarize common failure patterns, providing insights to guide future improvements in LLM-generated reasoning evaluation.

\noindent\textbf{Contributions.}
The main contributions of this paper are:
\begin{itemize}[leftmargin=15pt]
    \item \textbf{A post-hoc evaluation framework for LLM-generated reasoning.} We propose \toolname, a framework that integrates formal methods to evaluate LLM-generated reasoning by verifying both step-level correctness and overall logical consistency. \toolname is open-sourced at the repository~\cite{matp}.
    \item \textbf{A novel fine-grained reasoning chain classification.} We introduce a new classification scheme that distinguishes between sound, partially flawed, and invalid reasoning chains, enabling more nuanced evaluation of LLM-generated reasoning quality.
    \item \textbf{Analysis of LLM-generated reasoning behaviors.} By large-scale evaluation and classification, we uncover model-specific reasoning patterns and failure modes, and identify common NL2FOL translation errors that inform future improvements.
\end{itemize}
\vspace{-2mm}
\section{Background and Motivation}

\subsection{Logic Reasoning in LLMs}

Logical reasoning tasks performed by LLMs are generally categorized into \emph{inductive}, \emph{abductive}, and \emph{deductive} reasoning. Among these, deductive reasoning is particularly crucial and forms the core focus of our work. It emphasizes rigorous, consistent, and systematic inference from explicitly stated premises, making it essential in high-assurance domains such as formal verification, legal reasoning, and security-critical decision-making.
Recent neuro-symbolic efforts have shown the feasibility of transforming natural language into formal representations and LLM–Prover integration for formal reasoning~\cite{pan2023logiclm, olausson-etal-2023-linc, NEURIPS2023_8e9c7d4a}. 
Inspired by these efforts, we propose a lightweight divide-and-conquer framework for assessing logical validity in LLM-generated deductive reasoning chains.

\noindent\textbf{A motivating example.}
\autoref{fig:motivation} illustrates why systematic verification of LLM reasoning is necessary. When given a deductive reasoning task from \proofwriter{}, both GPT-3.5-Turbo and GPT-4o reach the correct conclusion but through vastly different reasoning processes. Notably, both models express their reasoning in natural language without an explicit logical structure, making it difficult to verify the correctness of individual steps and their interconnections. 
This highlights \textbf{Challenge \#1}. GPT-3.5-Turbo's reasoning contains critical flaws: steps 3 and 7 introduce unsupported information while step 13 contradicts the premises—yet these errors are masked by fluent natural language. This illustrates \textbf{Challenge \#2} for identifying subtle yet consequential reasoning errors. Moreover, even when individual steps seem correct, the entire reasoning chain generated by GPT-3.5-Turbo still fails to support the conclusion, indicating that the answer might be coincidental. 
\vspace{-2mm}
\subsection{Formal Foundations}

We outline the technical foundations for formalizing and verifying logical reasoning using first-order logic (FOL) and automated theorem provers (ATPs).

\noindent\textbf{FOL} serves as a formal language for representing and reasoning about logical statements.
It extends propositional logic with predicates and quantified variables, enabling the expression of general statements about objects in a domain.
The core components of FOL include: \emph{constants} representing specific objects, \emph{predicates} representing properties or relations, \emph{logical connectives} such as negation ($\neg$), conjunction ($\wedge$), disjunction ($\lor$), and implication ($\rightarrow$), and \emph{quantifiers} ($\forall$ for ``for all'', $\exists$ for ``there exists'').


The central criterion for evaluating such reasoning is \emph{logical validity}.
A conclusion $C$ is a logical consequence of a set of premises $P = \{P_1, P_2, \ldots, P_n\}$, denoted as $P\models C$, if and only if $C$ is true in every model (i.e., every possible interpretation of the symbols) that satisfies all premises in $P$.
This property of validity ensures that there can be no counterexample where the premises hold true but the conclusion does not, providing a rigorous standard for correctness.
In sound and complete logical systems like FOL, this semantic entailment ($\models$) is equivalent to syntactic provability ($\vdash$), meaning a valid conclusion can also be derived from the premises through a finite sequence of formal inference steps.
Our work aims to automatically check if the content generated by LLMs satisfy this standard of validity with respect to the given premises.

\noindent\textbf{ATPs} are specialized tools designed to automatically determine the validity of logical entailments in formal systems like FOL.
State-of-the-art ATPs, including Vampire~\cite{kovacs2013vampire} used in our work, primarily operate on the principle of proof by refutation.
To verify the logical validity $P\models C$, the prover mechanistically attempts to find a syntactic proof for the unsatisfiability of $P \cup \{\neg C\}$, i.e., it searches for a derivation $P \cup \{\neg C\} \vdash \bot$, where $\bot$ represents a contradiction.
Due to the completeness of FOL, finding such a proof provides a definitive confirmation of the original semantic entailment $P \models C$.
For interoperability, our framework translates FOL formulas into the standardized TPTP (Thousands of Problems for Theorem Provers) format~\cite{sutcliffe2017tptp}.
Using a high-performance ATP like Vampire enables the rigorous and automated validation of reasoning steps at scale.

\section{Methodology}
\label{method}

\begin{figure*}[ht]
    \vspace{-1mm}
    \centering
    \includegraphics[width=\textwidth]{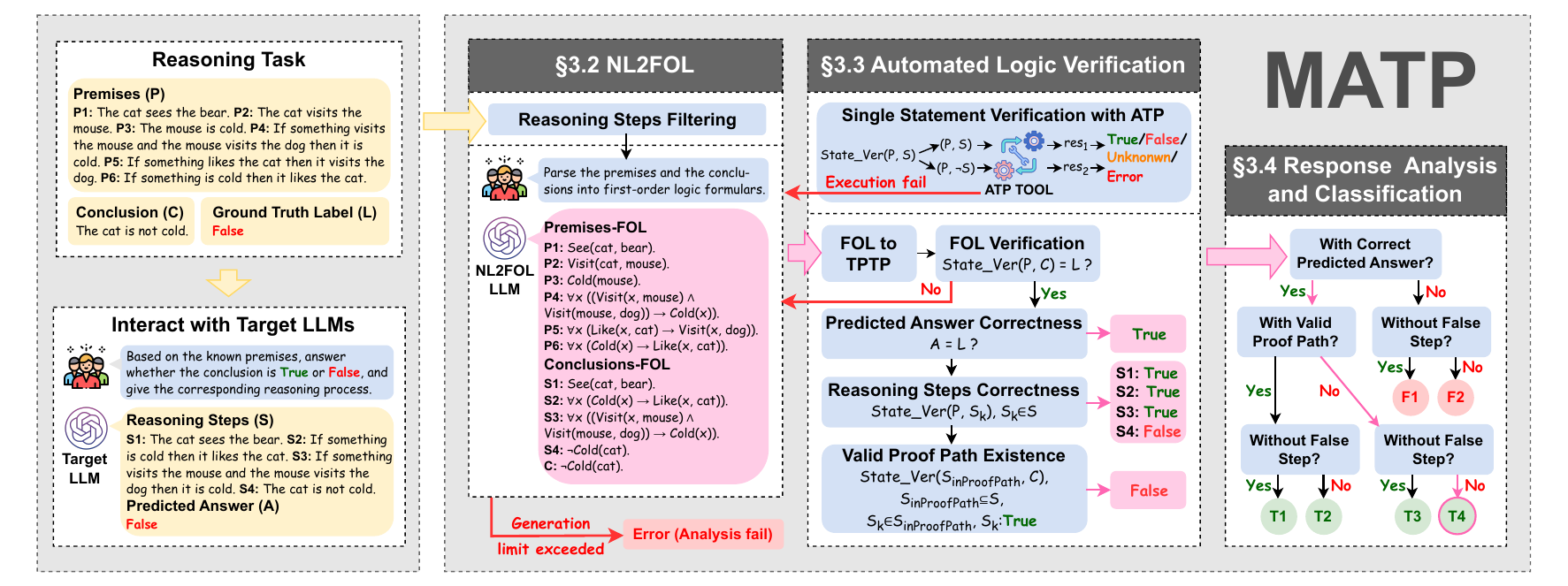}\vspace{-0.3cm}
    \caption{The workflow of \toolname{}. \textcolor{myyellow}{Yellow} nodes indicate the input, while \textcolor{mypink}{pink} nodes represent intermediate outputs.}\vspace{-0.3cm}
    \label{fig:overview}
\end{figure*}

We develop a fine-grained evaluation framework for LLM-generated reasoning in deductive tasks, focusing on more than just the final answer's correctness~(\autoref{sec:taskformalization}).
Specifically, \toolname{} uses an LLM to convert both reasoning tasks and model-generated responses into formal representations~(\autoref{sec:nl2fol}), employs an ATP tool and designed algorithms to verify step-level correctness and overall validity of the reasoning chain~(\autoref{sec:logical_verification}), and classifies the reasoning chain into fine-grained categories~(\autoref{sec:result_analysis}). An overview is shown in \autoref{fig:overview}.

\vspace{-1.5mm}
\subsection{Task Formalization}
\label{sec:taskformalization}

Given a set of premises and a candidate conclusion, the target LLM is prompted to determine whether the conclusion logically follows from the premises and to generate reasoning to support its judgment.
\toolname{} takes the problem statement and the generated reasoning content as input, and performs formal verification to assess both the correctness of individual reasoning steps and the overall sufficiency of the reasoning chain.
Formally, the input to \toolname{} is defined as $D=\{P,C,L,S,A\}$, consisting of five components:
\begin{enumerate}[leftmargin=15pt]
    \item \textbf{Premises~($P$):} A set of factual statements $P = \{P_1, P_2, \dots, P_n\}$, where each $P_i$ is assumed to be true for subsequent reasoning.
    \item \textbf{Candidate Conclusion~($C$):} A statement that is subject to validation based on the given premises $P$. By design, it is guaranteed to either logically follow from $P$ or contradict it.
    \item \textbf{Ground Truth Label~($L$):} A manually annotated label $L \in \{True, False\}$ representing the correctness of the candidate conclusion $C$ under the given premises $P$. Specifically, $True$ indicates that $P \vdash C$, i.e., $C$ logically follows from $P$, and $False$ indicates that $P \vdash \neg C$, i.e., $C$ contradicts $P$.
    \item \textbf{Reasoning Steps~($S$):} A sequence of intermediate statements $S = (S_1, S_2, \dots, S_m)$ generated by the target LLM during its reasoning process, where each $S_i$ represents a statement produced as part of the model's reasoning trajectory toward validating or refuting the conclusion.
    \item \textbf{Predicted Answer~($A$):} The final judgment $A \in \{True, False\}$ produced by the target LLM regarding the correctness of the candidate conclusion $C$, representing the model's decision based on the reasoning steps $S$.
\end{enumerate}

Based on the input $D$, \toolname{} translates the premises $P$, candidate conclusion $C$, and reasoning steps $S$ into FOL formulas and performs logical verification using an automated theorem prover. 
It produces three key outputs to facilitate comprehensive evaluation:

\begin{enumerate}[leftmargin=15pt]
    \item \textbf{Reasoning Step Correctness Annotation:} For each reasoning step, \toolname{} assigns a correctness label
    \begin{math}
        StepCorrectness(S_i)\newline \in \{True, False, Unknown\}
    \end{math},
    indicating whether $S_i$ is provable, refutable, or indeterminate based on the premises $P$.
    \item \textbf{Valid Proof Path Existence:} \toolname{} searches for a subset of correct reasoning steps $S_{inProofPath} \subseteq S$ that together entail the conclusion $C$ or its negation $\neg C$. If such a valid proof path exists, it indicates that the reasoning chain is logically sound.
    \item \textbf{Fine-Grained Classification of the Reasoning Chain}: By analyzing the correctness of the final predicted answer, the correctness of individual reasoning steps, and the existence of a valid proof path, \toolname{} categorizes the reasoning output into distinct fine-grained classes.
\end{enumerate}

Through this principled workflow, \toolname{} enables a rigorous and interpretable assessment of LLM reasoning capabilities, going beyond surface-level correctness to reveal potential logical errors and providing insights into the underlying reasoning structure.

\vspace{-2mm}
\subsection{NL2FOL}
\label{sec:nl2fol}

This module converts the given premises $P$, candidate conclusion $C$, and reasoning steps $S$ into FOL representations to enable verification with automated theorem proving.

\subsubsection{Preprocessing.} To ensure the accuracy and reliability of translating NL to FOL, we first apply a series of preprocessing measures. 

\noindent\textbf{Interacting with LLMs.}
To generate reasoning chains for subsequent formalization and verification, we employ a structured interaction with LLMs.
Given premises $P$ and a candidate conclusion $C$, the model is prompted to assess whether $C$ logically follows from $P$ through an explicitly generated reasoning process.
We use a few-shot prompting strategy, providing annotated examples to guide the model's behavior. 
The LLM is instructed to first produce a step-by-step reasoning process based solely on $P$, followed by a \textit{True/False} decision on whether $C$ is entailed.
This answer-after-reasoning format decouples the reasoning trace from the final verdict, enabling fine-grained evaluation of both components. To ensure faithful reasoning, the prompt explicitly constrains the output to avoid unnecessary complexity, ambiguous language, or reliance on external knowledge. The full prompt design is included in ~\cite{matp}.

\noindent 
\textbf{Further filtering.}
After generating the reasoning process, the output is split into individual sentences.
Subsequently, uncertain or speculative statements~(e.g., ``Wren is not necessarily a gorpus'') are filtered out. 
Inspired by prior work~\cite{cui2025practical}, which segments reasoning content using reflection-related keywords to identify reflective phases, we extend this approach.
We remove uncertain or speculative statements by detecting ambiguity-indicative expressions (e.g., ``possible'', ``contradict'', or ``not necessarily''), without introducing additional filtering models.
The remaining reasoning steps are concise and explicit, aiding downstream processing.
In particular, repetitive statements are deduplicated after converting sentences into FOL form, as direct comparison in natural language is difficult.

\begin{figure}[t!]
\resizebox{0.48\textwidth}{!}{
\begin{tcolorbox}[
enhanced, 
boxrule=1pt,
title=Prompt Template for Translating NL to FOL,
left=2pt, right=2pt, top=2pt, bottom=2pt,  
boxsep=2pt,fontupper=\scriptsize
]

Parse the given premises and conclusions into first-order logic formulas. 
\\
\textcolor[HTML]{4169E1}{\textbf{\#\#\# Rules}} \\
1. Each sentence must be converted into a single FOL formula. \\
2. Use the following FOL syntax: logical conjunction of expr1 and expr2: $expr1 \land expr2$ ... \\
3. Every formula must explicitly include quantifiers ($\forall$x or $\exists$x) where applicable ... \\
4. When an existential entity reappears with new properties, all properties (both old and new) should be conjoined within the scope of a single existential quantifier into the later formula ...
\\
... (other more specific rules)
\\
\textcolor[HTML]{4169E1}{\textbf{\#\#\# Output Format}} \\
Premises / Conclusions: fol-formula ::: (index) reference. ...
\\
\textcolor[HTML]{4169E1}{\textbf{\#\#\# Examples}} \\
Premises: (1) The cow is cold. (2) The cow is not round. ... \\
Conclusions: (1) The mouse needs the cow. ... (9) The mouse eats the mouse.\\
\#\#\#\\
Premises: Cold(cow) ::: (1) The cow is cold. ¬Round(cow) ::: (2) The cow is not round. ... \\
Conclusions: Needs(mouse, cow) ::: (1) The mouse needs the cow. ... Eats(mouse, mouse) ::: (9) The mouse eats the mouse.
\\
\textcolor[HTML]{4169E1}{\textbf{\#\#\# Your Task}} \\
Return first-order logic formulas with NO other texts. \\
Premises: \textcolor{blue}{\{premises\}}\\
Conclusions: \textcolor{blue}{\{conclusions\}}
\end{tcolorbox}}
\vspace{-0.6cm}
\caption{Prompt Template for Translating NL to FOL.}
\vspace{-0.6cm}
    \label{fig:nlfol_prompt}
\end{figure}

\subsubsection{Translating Natural Language into First-Order Logic.} 
Inspired by Logic-LM~\cite{pan2023logiclm},
which successfully leverages prompt engineering with GPT-4 for formalizing long-text reasoning tasks, we follow its prompt design framework to implement NL2FOL.
While retaining its core structure, we introduce several modifications tailored to the specific requirements of our task. 

\noindent\textbf{Prompt Structure.}
As illustrated in \autoref{fig:nlfol_prompt}, the prompt structure comprises five components: \emph{task definition}, \emph{translation rules}, \emph{output format}, \emph{example demonstrations}, and \emph{input instance}, detailed as follows. 
Specifically, the ``\texttt{conclusions}'' in the prompt always include both the reasoning steps~$S$ and the candidate conclusion~$C$ from the actual input.

\emph{Task definition} establishes the primary goal of converting premises and conclusions into corresponding FOL representations. 
\emph{Translation rules} define the fundamental constraints that govern the NL2FOL process.
They ensure that each statement is translated in strict accordance with FOL syntax, while following a set of quality assurance strategies.
These rules can be flexibly adjusted based on specific scenarios when necessary.
\emph{Output format} requires that each generated output include a formal FOL expression, an index, and the corresponding reference sentence from the input.
To ensure completeness and avoid omissions, both premises and conclusions are numerically indexed, accommodating potentially lengthy inputs.
\emph{Example demonstrations} show the expected input-output relationships, enhancing clarity and reinforcing the correct application of conversion principles.
\emph{Input instance} provides the reasoning input, where ``\texttt{Premises}'' corresponds to premises~$P$, and ``\texttt{Conclusions}'' includes both reasoning steps~$S$ and the candidate conclusion~$C$. 
The model is instructed to output only the required FOL formulas, without any extraneous content.

\noindent \textbf{Quality Assurance Strategies.}
In the \emph{Rules} component, we implement strategies to minimize common errors and improve translation quality.
For \textbf{quantifier usage}, variables must have explicit existential or universal quantifiers. Parentheses are applied to complex statements to define quantifier scopes and avoid ambiguity.
For existentially quantified entities, multiple statements can describe different attributes of the same entity, such as introducing property \emph{A} for an entity \emph{Tom} and later adding property \emph{B}.
The later translation should integrate both properties to preserve semantic completeness.
To ensure \textbf{cross-step coherence in existential instantiation}, we enforce rule constraints to consistently reference entities and their predicate attributes introduced in earlier steps, avoiding logical inconsistencies.
As we translate premises, reasoning steps, and candidate conclusions in a single pass, LLMs’ comprehension aids predicate consistency. 
To further ensure \textbf{consistent predicate usage}, we add the rule that encourages reusing existing predicates before introducing new ones. This reduces redundancy and prevents predicate proliferation.
To address \textbf{LLM's confusion} between ``exclusive or'' and ``inclusive or'', we include targeted conversion examples to clarify their distinction.
Detailed implementations of these strategies are provided in the full prompt~\cite{matp}.

\subsubsection{Regeneration}
The translation from natural language to FOL by LLMs is not always correct initially.
If the generated FOL results in execution errors during verification, or if the verification outcome of the candidate conclusion contradicts the ground-truth label, the NL2FOL process is retried up to a fixed maximum number of attempts. If all retries fail, an error is returned, indicating failure to analyze the input.
No feedback is provided during regeneration. Since pinpointing semantic errors in translation is challenging, we cannot provide useful and specific feedback. While using the entire output from the previous generation as feedback is another option, experiments in~\autoref{sec:AblationStudy} show its minimal impact.

\subsection{Logical Verification}
\label{sec:logical_verification}

This module aims to perform logical verification on FOL representations of input statements ($P$, $C$, $S$) using the automated theorem prover. This process assigns correctness labels to each reasoning step and determines whether a valid proof path exists.

\noindent \textbf{FOL to TPTP.} 
We convert FOL formulas into TPTP format for compatibility with standard ATP tools. This is done via a Python script that enforces a deterministic one-to-one mapping between FOL and TPTP representations. 
Preprocessing steps are applied to align the FOL formulas with TPTP syntax, including replacing Unicode symbols with ASCII characters, normalizing predicate and constant names, and handling unsupported logical connectives (e.g., exclusive disjunction $A \oplus B$ is rewritten as $(\neg A \land B) \lor (A \land \neg B)$). The script also performs lightweight syntax repair—such as correcting missing parentheses or misused connectives—to improve robustness and reduce parsing errors in downstream verification.


Each formula is translated into TPTP’s FOF format: ``\texttt{fof}(\emph{name}, \emph{role}, \emph{formula})'', where \emph{name} is a unique identifier for the logical statement, \emph{role} specifies whether it is an axiom or a conjecture, and \emph{formula} is the FOL expression rewritten in TPTP syntax.
All premises are assigned as axioms, and the candidate conclusion as a conjecture.
Unique identifiers follow the pattern ``\emph{premise\_index}'' and ``\emph{conclusion}''.
We first translate natural language into FOL before TPTP because TPTP syntax is more complex and verbose, making direct NL2TPTP translation more error-prone. 
Converting FOL to TPTP helps correct syntax errors and ensures compatibility with theorem provers, improving verification reliability.

    \noindent\textbf{Provability Verification.}
    A core component of the verification process is to determine whether a statement $S$ is provable from a set of premises $P$. 
    There are three possible outcomes for $S$: \emph{provable}, which means $S$ is logically entailed by $P$ and it holds true whenever $P$ is true; \emph{refutable}, which means $S$ contradicts $P$ and is always false when $P$ is true; \emph{indeterminate}, which means $S$ is neither provable nor refutable based on $P$.
    To determine this, we invoke the prover to simultaneously check whether $P \vdash S$ or $P \vdash \neg S$. If either holds, $S$ is classified as \emph{True} or \emph{False} accordingly. If neither holds (i.e., both $P \cup {S}$ and $P \cup {\neg S}$ are consistent), $S$ is labeled \emph{Unknown}.
    \autoref{alg:single_step_verification} summarizes this procedure, with the outputs \emph{True}, \emph{False}, and \emph{Unknown} corresponding to the three provability outcomes.
    In particular, if both $S$ and $\neg S$ are provable, it indicates a contradiction within $P$, and we return an \emph{Error}.

\vspace{-2mm}
\begin{algorithm}
\scriptsize
\caption{Single Statement Verification using ATP}
\label{alg:single_step_verification}
\begin{algorithmic}[1]
\Require Premises $P = \{ P_1, P_2, \dots, P_n \}$, candidate statement $S$
\Ensure  Check Result (\textbf{True}, \textbf{False}, \textbf{Unknown}, or \textbf{Error})
\Function{VerifySingleStatement}{$P, S$}
    \State $res_1 \gets $\Call{ATPRun}{$P, S$} \textbf{and} $res_2 \gets $\Call{ATPRun}{$P, \neg S$} 
    \If{ \textbf{Refutation} \textbf{in} $res_1$ \textbf{and}  \textbf{Refutation} \textbf{in} $res_2$}
        \State \Return \textbf{Error} \Comment{\commentstyle{Contradiction exists in $P$}}
    \EndIf
    \If{\textbf{Refutation} \textbf{in} $res_1$} \State \Return \textbf{True} \Comment{\commentstyle{$S$ is verified}}
    \EndIf
    \If{\textbf{Refutation} \textbf{in} $res_2$} \State \Return \textbf{False} \Comment{\commentstyle{$\neg S$ is verified}}
    \EndIf
    \State \Return \textbf{Unknown}
\EndFunction
\end{algorithmic}
\end{algorithm}
\vspace{-2mm}

    \noindent\textbf{Verification of the Reasoning Steps.}
    We begin by applying \textsc{VerifySingleStatement} to the premises $P$ and the candidate conclusion $C$ to check alignment with the ground-truth label $L$. 
    A mismatch may arise from: (1) a semantically incorrect but syntactically valid NL2FOL translation that causes the ATP to produce a judgment on $C$ inconsistent with $L$, or (2) an erroneous ground-truth label annotation.
    Since these cases cannot be automatically differentiated or corrected, the process reverts to the NL2FOL stage for regeneration.
    After verifying the conclusion, we assess the correctness of each intermediate reasoning step $S_i$ under the same premises $P$. If a step is refuted or indeterminate, it is recorded as a potential logical error.
    \autoref{alg:reasoning_steps_verification} outlines this process. The algorithm returns both the correctness of the predicted answer and the correctness annotation for each reasoning step. Notably, some incorrect steps may not affect the final judgment, while others may compromise the entire reasoning chain. To distinguish these cases, further analysis is conducted to determine whether a valid proof path exists within the reasoning steps.

\vspace{-2mm}
\begin{algorithm}
\scriptsize
\caption{Reasoning Steps Verification}
\label{alg:reasoning_steps_verification}
\begin{algorithmic}[1]

\Require Premises $P = \{ P_1, P_2, \dots, P_n \}$, candidate conclusion $C$, ground truth label $L$, reasoning steps $S = \{ S_1, S_2, \dots, S_m \}$, predicted answer $A$
\Ensure Returns $AnswerCorrectness$, step correctness annotation $S_{correctness}$

\Function{VerifyReasoningSteps}{$P, C, L, S, A$} 

    \If{\Call{VerifySingleStatement}{$P, C$} $\ne$ $L$}
        \State \Return \textbf{Error} \Comment{\commentstyle{NL2FOL translation error or labeling error}}
    \EndIf
    
    \State $S_{correctness} \gets []$ \Comment{\commentstyle{Initialize step correctness list}}
    \For{$i \gets 1$ to $m$}
        \State \textsc{Append}($S_{correctness}$, \Call{VerifySingleStatement}{$P, S_i$})
    \EndFor

    \State $AnswerCorrectness \gets (L = A)$ \Comment{\commentstyle{Check if the answer matches}}
    
    \State \Return $AnswerCorrectness$, $S_{correctness}$ 

\EndFunction

\end{algorithmic} 
\end{algorithm}
\vspace{-2mm}

\noindent \textbf{Verification of the Reasoning Proof Path.} 
Beyond evaluating individual reasoning steps, we also assess whether the reasoning process contains all the necessary steps to logically support the conclusion label. In other words, we determine whether the reasoning chain includes a valid proof path leading to the correct classification of the conclusion.
    To do this, we iteratively construct a candidate proof path by including only those \emph{True}-labeled steps that introduce new facts not derivable from previously included steps. Formally, the proof path is a subset $S_{inProofPath} \subseteq S$, where a step $S_i$ is added if and only if it is labeled \emph{True} and not entailed by $S_{inProofPath}$.
    After processing all steps, we check if the truth value of the candidate conclusion $C$ matches the ground truth label $L$, when evaluated under the proof path $S_{inProofPath}$. 
    If True, it indicates that reasoning process contains a valid proof path that supports the conclusion label.
    Otherwise, the proof path is considered invalid.
    This verification process is detailed in \autoref{alg:reasoning_path_verification}.

\vspace{-2mm}
\begin{algorithm}
\scriptsize
\caption{Reasoning Proof Path Verification}
\label{alg:reasoning_path_verification}
\begin{algorithmic}[1]

\Require Candidate conclusion $C$, ground truth label $L$, reasoning steps $S = \{ S_1, S_2, \dots, S_m \}$, step correctness annotation $S_{correctness}$
\Ensure Returns $HasValidProofPath$

\Function{VerifyReasoningProofPath}{$C, L, S, S_{correctness}$}

    \State $S_{inProofPath} \gets []$
    \For{$i \gets 1$ to $m$}
        \If{$S_{correctness\_i} = \textbf{True}$}
            \If{\Call{VerifySingleStatement}{$S_{inProofPath}, S_i$} $= \textbf{Unknown}$}
                \State \textsc{Append}($S_{inProofPath}, S_i$) \Comment{\commentstyle{Construct the proof path}}
            \EndIf
        \EndIf
    \EndFor
    \State $HasValidProofPath \gets ($\Call{VerifySingleStatement}{$S_{inProofPath}, C$}~$= L)$
    \State \Return $HasValidProofPath$
\EndFunction

\end{algorithmic}
\end{algorithm}
\vspace{-2mm}

If the automated theorem prover encounters an execution error indicating syntactic flaws in the generated FOL content, we return to the NL2FOL stage for regeneration.


\subsection{Response Analysis and Classification}
\label{sec:result_analysis}
After the first two stages, inputs exceeding the maximum NL2FOL retries are categorized as \textbf{Error}, indicating that our framework cannot handle them. 
For successfully processed inputs,
this module performs a fine-grained classification of reasoning chains based on logical verification to evaluate the reasoning process comprehensively.
Instead of only assessing the final predicted answer, we analyze both individual reasoning steps and the overall reasoning chain to identify potential logical flaws. 
This allows us to distinguish between responses that are correct by chance and those based on valid reasoning.
After logical verification, we obtain three key evaluation dimensions: (1) correctness of the predicted answer, (2) correctness of each reasoning step, and (3) existence of a valid proof path within the reasoning chain. Based on these dimensions, we define six reasoning categories, as illustrated in \autoref{fig:overview}. 

For responses with incorrect predicted answer, we classify them into two categories based on whether intermediate reasoning steps contain errors. Any reasoning step labeled as $False$ or $Unknown$ is considered erroneous. 
\textbf{F1} represents cases where all reasoning steps are correct but the final answer is incorrect, suggesting that the target model likely fails to construct a valid proof path. 
\textbf{F2} covers cases where both the final answer and at least one reasoning step are incorrect, suggesting that intermediate errors may mislead the reasoning process or that the model fails to form a coherent proof.

For responses with correct predicted answers, we assess both the correctness of individual steps and the validity of the reasoning chain.
\textbf{T1} denotes the ideal cases where all steps are correct and a valid proof path exists. 
\textbf{T2} includes cases with some erroneous steps, yet the proof path remains valid—indicating partial robustness.
\textbf{T3} represents cases where all reasoning steps are correct, but the reasoning chain does not constitute a valid proof, suggesting that the model correctly infers the answer but fails to provide a logically complete reasoning path. 
\textbf{T4} includes responses with step-level errors and no valid proof path, raising concerns that the correct answer may be reached by chance or flawed reasoning.

\section{Evaluation}
\label{sec:eval}

Our evaluation targets the following research questions:

\begin{itemize}[leftmargin=15pt]
    \item \textbf{RQ1~(Effectiveness): }Can \toolname{} accurately verify the correctness of reasoning steps generated by LLMs?  
    \item \textbf{RQ2~(Reasoning Chain Classification): }Can \toolname{} effectively classify different types of reasoning chains? 
    \item \textbf{RQ3~(Model Reasoning Analysis): }How do different LLMs perform in terms of reasoning quality, and what are their characteristic behaviors when generating reasoning chains?
\end{itemize}

\begin{table*}[!htbp]
\centering
\fontsize{9}{12}\selectfont
\caption{Evaluation of \toolname{} on the Benchmark: NL2FOL Translation Accuracy and Reasoning Step Correctness Classification Performance. NL2FOL accuracy~(\autoref{sec:rq1_1}) is evaluated using four metrics—ER (Execution Rate), EA (Execution Accuracy), FB (FOL BLEU), and LE (Logical Equivalence)—while reasoning step correctness~(\autoref{sec:rq1_2}) is measured by Macro F1 score. All values are reported as percentages (\%), with higher scores indicating better performance. In the metric rows, GPT-4o and DS-R1 refer to prompt-based baseline methods implemented using GPT-4o and DeepSeek-R1, respectively. 
}\vspace{-0.3cm}
\label{table:rq1_1}
\resizebox{\textwidth}{!}{
\begin{tabular}{c|ccccccc|ccccccc|ccccccc}
\hline 
\multicolumn{1}{c|}{\multirow{3}{*} {
\diagbox[width=9em,height=3.9em]{\textbf{Target LLM}}{\textbf{Dataset}}}} & \multicolumn{7}{c|}{\textbf{\prontoqa{}~(384)}} & \multicolumn{7}{c|}{\textbf{\proofwriter{}~(344)}} & \multicolumn{7}{c}{\textbf{\folio{}~(355)}}\\
\cline{2-22}
\multicolumn{1}{c|}{} & 
\multicolumn{1}{c}{\multirow{2}{*} {\textbf{ER}}} & \multicolumn{1}{c}{\multirow{2}{*} {\textbf{EA}}} & \multicolumn{1}{c}{\multirow{2}{*} {\textbf{FB}}} & \multicolumn{1}{c}{\multirow{2}{*} {\textbf{LE}}}  & \multicolumn{3}{c|}{\textbf{Macro F1 Score}} &
\multicolumn{1}{c}{\multirow{2}{*} {\textbf{ER}}} & \multicolumn{1}{c}{\multirow{2}{*} {\textbf{EA}}} & \multicolumn{1}{c}{\multirow{2}{*} {\textbf{FB}}} & \multicolumn{1}{c}{\multirow{2}{*} {\textbf{LE}}}  & \multicolumn{3}{c|}{\textbf{Macro F1 Score}} &
\multicolumn{1}{c}{\multirow{2}{*} {\textbf{ER}}} & \multicolumn{1}{c}{\multirow{2}{*} {\textbf{EA}}} & \multicolumn{1}{c}{\multirow{2}{*} {\textbf{FB}}} & \multicolumn{1}{c}{\multirow{2}{*} {\textbf{LE}}}  & \multicolumn{3}{c}{\textbf{Macro F1 Score}}  \\
\multicolumn{1}{c|}{} & 
\multicolumn{1}{c}{} & \multicolumn{1}{c}{} & \multicolumn{1}{c}{} & \multicolumn{1}{c}{}  & \textbf{MATP} & \textbf{GPT-4o} & \textbf{DS-R1} &
\multicolumn{1}{c}{} & \multicolumn{1}{c}{} & \multicolumn{1}{c}{} & \multicolumn{1}{c}{}  & \textbf{MATP} & \textbf{GPT-4o} & \textbf{DS-R1} &
\multicolumn{1}{c}{} & \multicolumn{1}{c}{} & \multicolumn{1}{c}{} & \multicolumn{1}{c}{}  & \textbf{MATP} & \textbf{GPT-4o} & \textbf{DS-R1} \\
\hline
\multicolumn{1}{c|}{\textbf{GPT-3.5}} 
& 99.22 & 100.0 & 100.0 & 100.0 & 87.13 & 54.33 & 39.94
& 99.71 & 99.71 & 83.88 & 99.91 & 100.0 & 43.03 & 38.39 
& 96.34 & 69.59 & 36.98 & 88.37 & 91.93 & 57.11 & 43.21 \\
\multicolumn{1}{c|}{\textbf{GPT-4o}} 
& 98.96 & 100.0 & 100.0 & 100.0 & 100.0 & 49.75 & 47.92 
& 99.71 & 99.42 & 84.14 & 100.0 & 100.0 & 33.08 & 33.33 
& 96.06 & 70.09 & 37.11 & 87.78 & 100.0 & 42.23 & 31.06 \\
\multicolumn{1}{c|}{\textbf{LLaMA3.1-8B}} 
& 99.22 & 100.0 & 100.0 & 100.0 & 100.0 & 38.53 & 49.74 
& 99.71 & 100.0 & 83.88 & 99.91 & 100.0 & 46.77 & 40.00 
& 95.49 & 70.50 & 36.84 & 88.52 & 89.87 & 68.01 & 47.08 \\
\multicolumn{1}{c|}{\textbf{LLaMA3.1-70B}} 
& 99.22 & 99.74 & 100.0 & 100.0 & 55.42 & 49.59 & 69.69 
& 99.71 & 97.38 & 83.69 & 99.91 & 98.61 & 46.10 & 43.05 
& 94.37 & 71.64 & 36.46 & 88.20 & 93.51 & 38.53 & 42.12 \\
\multicolumn{1}{c|}{\textbf{Qwen2.5-7B}} 
& 98.96 & 100.0 & 100.0 & 100.0 & 100.0 & 44.43 & 32.80  
& 99.42 & 99.42 & 84.14 & 100.0 & 96.34 & 52.80 & 47.29 
& 97.18 & 72.46 & 36.35 & 88.15 & 79.69 & 49.34 & 38.40  \\
\multicolumn{1}{c|}{\textbf{QwQ-32B}} 
& 99.22 & 100.0 & 100.0 & 100.0 & 100.0 & 49.37 & 49.79  
& 99.13 & 99.41 & 84.28 & 100.0 & 100.0 & 72.70 & 49.27  
& 95.49 & 71.09 & 36.63 & 88.16 & 88.75 & 37.29 & 36.55  \\
\multicolumn{1}{c|}{\textbf{DS-R1-Qwen7B}} 
& 99.22 & 99.74 & 100.0 & 100.0 & 100.0 & 44.27 & 39.96  
& 99.42 & 98.25 & 84.14 & 100.0 & 98.84 & 62.44 & 36.66  
& 96.34 & 70.76 & 37.81 & 88.65 & 72.08 & 41.72 & 39.66  \\
\multicolumn{1}{c|}{\textbf{DS-R1-Qwen32B}} 
& 99.22 & 99.74 & 100.0 & 100.0 & 100.0 & 49.58 & 83.13  
& 99.42 & 99.42 & 84.28 & 100.0 & 66.47 & 32.52 & 32.42  
& 96.62 & 71.43 & 37.29 & 88.92 & 72.82 & 38.53 & 33.70  \\
\multicolumn{1}{c|}{\textbf{DeepSeek-R1}} 
& 99.22 & 99.74 & 100.0 & 100.0 & 100.0 & 49.82 & 100.0 
& 99.42 & 99.42 & 84.21 & 100.0 & 52.16 & 49.19 & 49.35 
& 97.18 & 70.72 & 37.06 & 87.82 & 61.33 & 32.82 & 31.33 \\
\multicolumn{1}{c|}{\textbf{GPT-o4-mini}} 
& 98.18 & 100.0 & 100.0 & 100.0 & 100.0 & 48.19 & 100.0  
& 99.71 & 100.0 & 84.07 & 100.0 & 100.0 & 54.96 & 54.49  
& 95.49 & 70.50 & 36.58 & 88.45 & 94.24 & 46.13 & 32.40  \\
\hline
\multicolumn{1}{c|}{\textbf{Average}} 
& 99.06 & 99.90 & 100.0 & 100.0 & 94.26 & 47.79 & 61.30
& 99.53 & 99.24 & 84.07 & 99.97 & 91.24 & 49.36 & 42.43
& 96.06 & 70.88 & 36.91 & 88.30 & 84.42 & 45.17 & 37.55
\\
\hline
\end{tabular}}\vspace{-0.3cm}
\end{table*}

\vspace{-2mm}
\subsection{Experimental Setup}
\noindent \textbf{Benchmark Dataset.} 
We construct our benchmark from three reasoning-focused datasets, each designed for natural language inference tasks.
\prontoqa{}~\cite{PrOntoQAOOD} and \proofwriter{}~\cite{tafjord2021proofwriter} focus on logical reasoning within structured ontologies, emphasizing categories, attributes, and rule-based inference, while \folio{}~\cite{han2024folio} centers on realistic semantic reasoning with knowledge about people, locations, quantities, and events. 
Each test case in these datasets consists of premises, a candidate conclusion, and a ground-truth label ($True$, $False$, or $Unknown$).
Test cases in \prontoqa{} and \proofwriter{} use structured natural language, while those in \folio{} use expressions closer to everyday human language.

\begin{itemize}[leftmargin=15pt]
    \item \textbf{\prontoqa{}}~\cite{PrOntoQAOOD} (384 test cases selected) is a synthetic dataset where all conclusions are labeled $True$. Tasks are categorized by the minimal number of reasoning steps required; we sample from the 4-hop subset to ensure higher complexity.
    \item \textbf{\proofwriter{}}~\cite{tafjord2021proofwriter} (344 test cases selected) uses more natural language and complex logic. We sample from its 5-hop test set, excluding tasks with $Unknown$ label due to unclear entailment.
    \item \textbf{\folio{}}~\cite{han2024folio} (355 test cases selected) is a human-authored dataset grounded in real-world facts. We sample from the training set, retaining only $True$/$False$ cases and discarding instances where the conclusion trivially repeats a premise.
\end{itemize}

To balance dataset scale and focus on higher-hop instances that elicit richer model responses, we randomly sample test cases from the maximum-hop subset of \prontoqa{} and \proofwriter{}, rather than using all available data.
For \folio{}, which lacks hop-based partitioning, we randomly sample from its training set.
For each selected test case, we generate reasoning chains and predicted answers using 10 representative LLMs (detailed in the next section). The prompting template used for generation is provided in ~\cite{matp}.
This results in a total of 10830 benchmark instances. Each instance includes the premises, candidate conclusion and ground-truth label provided by the test case, as well as the reasoning chain and predicted answer generated by the target LLM.
This benchmark is used to evaluate the effectiveness of \toolname{} (RQ1) and to analyze reasoning chain categories across models (RQ3).

\noindent \textbf{Models Under Test.} To ensure a comprehensive evaluation, we test 10 LLMs with \toolname{}, including five general models and five reasoning models. The general models include GPT-3.5-turbo, GPT-4o~\cite{achiam2023gpt, hurst2024gpt}, LLaMA3.1-8B-chat-hf, LLaMA3.1-70B-chat-hf~\cite{grattafiori2024llama} and Qwen2.5-7B-Instruct~\cite{qwen2.5}.  The reasoning models include QwQ-32B~\cite{qwen32b}, DeepSeek-R1-Distill-Qwen-7B, DeepSeek-R1-Distill-Qwen-32B\footnote{We use DS-R1-Qwen7B/32B to represent DeepSeek-R1-Distill-Qwen-7B/32B for short.}, DeepSeek-R1~\cite{guo2025deepseek} and GPT-o4-mini~\cite{o4mini}. 

\noindent \textbf{Model Configuration.} 
We set \textit{temperature} to 0 for stable and consistent reasoning outputs, with \textit{top-p} = 1 and \textit{top-k} = 0 to favor high-probability tokens. An exception is GPT-o4-mini, which uses a fixed temperature of 1 due to model limitations.

\noindent \textbf{Baseline.} 
We compare our method against a prompting-based baseline using GPT-4o and DeepSeek-R1 (RQ1 and RQ2), following ~\cite{zeng2024mrgsm8kmetareasoningbenchmarklarge}. The prompt elicits two outputs: (1) step-by-step judgments on reasoning steps ($True$, $False$, or $Unknown$); and (2) a final decision on whether the reasoning supports the predicted answer ($Yes$ or $No$), both with explanations. The full prompt is provided in ~\cite{matp}.

\noindent \textbf{Implementation.}
Based on MetaGPT~\cite{hong2024metagpt}, a development framework for multi-agent systems that coordinates tool calls and manages pipeline stages, we integrate existing components to implement the evaluation framework as an automated toolchain.
Specifically, we use GPT-4o for the NL2FOL translation, leveraging its strong performance on long and complex natural language tasks, and limit generation attempts to three per input. For automated verification, we adopt the widely used automated theorem prover \vampire{}~\cite{kovacs2013vampire}. All implementation details are available at \cite{matp}.

\noindent \textbf{Running Environment.} 
All experiments run on a server with Ubuntu 22.04, equipped with two 64-core AMD EPYC 7713 CPUs, 512 GB RAM, and two NVIDIA A100 PCIe 80GB GPUs. 

\subsection{RQ1:~Effectiveness}
We conduct evaluations on the benchmark dataset to assess the effectiveness of \toolname{}. The evaluation focuses on two aspects: the accuracy of NL2FOL translation and the ability to identify the correctness of individual reasoning steps.

\subsubsection{Accuracy of NL2FOL translation.}
\label{sec:rq1_1}
We evaluate NL2FOL translation at both the overall and sentence levels. 

\noindent \textbf{Overall Level Metrics.} 
Following Logic-LM~\cite{pan2023logiclm}, we use Execution Rate~(ER) and Execution Accuracy~(EA) to evaluate NL2FOL translation across the entire benchmark. ER measures the proportion of FOL outputs that can be successfully executed by \vampire{}, reflecting syntactic correctness. EA measures how often \vampire{}'s verdict matches the ground truth label when execution succeeds, indicating semantic preservation. 

\noindent \textbf{Sentence Level Metrics.} 
We adopt FOL BLEU~(FB) and Logical Equivalence~(LE) from LogicLLaMA~\cite{yang2023logicllama} to assess sentence-level conversion. 
FB applies FOL-specific tokenization to compute similarity scores, while LE determines whether the generated and reference FOL formulas are logically equivalent.
For datasets with reference FOLs (\folio{}), we compute FB and LE across all premises. For others (\prontoqa{}, \proofwriter{}), we manually annotate reference FOLs for 300 premises per dataset for metric evaluation, with cross-verification by three authors to ensure consistency.

\noindent \textbf{Results.}
As shown in \autoref{table:rq1_1}, both ER and EA on \prontoqa{} and \proofwriter{} are close to 100\%, indicating strong syntactic and semantic alignment. In contrast, \folio{} shows high ER (96.06\%) but lower EA (70.88\%), suggesting challenges in preserving semantics for complex statements.
At the sentence level, FB and LE scores on \prontoqa{} and LE on \proofwriter{} are near 100\%. However, \proofwriter{}'s FB drops to 84.07\%, mainly due to mismatches in forms (e.g., “Like” vs. “Likes”) between model outputs and base-form annotations, highlighting FB's sensitivity to surface variations. On \folio{}, the FB score drops to 36.91\%, while LE remains relatively high (88.30\%), reflecting logical equivalence despite low textual similarity. This discrepancy is understandable, as for complex sentences, the choice of predicate or constant names is unlikely to consistently match the expected form.

To better understand the low EA on \folio{}, we conducted a manual review of cases where the \toolname{}'s labels for candidate conclusion differ from the ground truth labels. The key factors contributing to this issue will be analyzed in detail in \autoref{sec:rq3_3}.

\subsubsection{Ability to identify the correctness of individual reasoning steps.} 
\label{sec:rq1_2}

This evaluation focuses on whether each reasoning step is correctly labeled by \toolname{}, and how it performs versus the baseline. 

\noindent \textbf{Metric.} We formulate the task as a three-class classification problem ($True$, $False$, $Unknown$), evaluated using macro F1-score. Given the high annotation cost, we sample 20 ER-contributing tasks per dataset for each of the ten models, totaling 600 samples. Each reasoning step correctness within these samples is manually labeled by three authors, and the aggregated annotations are used to compute the final macro F1 score.

\noindent \textbf{Results.} The detailed results are shown in~\autoref{table:rq1_1}. \toolname{} consistently outperforms both GPT-4o and DeepSeek-R1 baselines in classifying reasoning step correctness. On \prontoqa{} and \proofwriter{}, \toolname{} achieves 94.26\% and 91.24\% F1, significantly higher than GPT-4o (47.79\%, 49.36\%) and DeepSeek-R1 (61.30\%, 42.43\%). Even on the more informal \folio{} dataset, \toolname{} maintains 84.42\% F1, while the baselines drop below 46\%. 
We further analyze the reasons behind the poor performance of the baselines. Specifically, the baselines often mislabel correct steps as $False$ or $Unknown$ when the inference is implicit or requires multi-hop reasoning. They also struggle to distinguish between $False$ and $Unknown$, resulting in frequent confusion between the two. 


\vspace{-1mm}
\begin{tcolorbox}[title=ANSWER to RQ1, boxrule=0.8pt,boxsep=1.5pt,left=2pt,right=2pt,top=2pt,bottom=1pt,fontupper=\small]
\toolname{} verifies reasoning steps by accurately translating NL to FOL and identifying correctness, outperforming prompting-based baselines across all datasets.  It performs well on \prontoqa{} and \proofwriter{}, but struggles on \folio{} due to more complex language patterns.
\end{tcolorbox} 
\vspace{-2mm}

\subsection{RQ2:~Reasoning Chain Classification} 
To evaluate whether \toolname{} can accurately classify reasoning chains, we conduct evaluations on a classification dataset and compare its performance with baselines. 

\noindent \textbf{Classification Dataset.}
We construct a classification dataset from 200 \prontoqa{} tasks. Starting from the standard reasoning chains provided by \prontoqa{} (T1-type), we apply targeted mutations to obtain different types of reasoning chains: to generate T2-type chains, we insert a negated premise to introduce false reasoning step; for T3-type chains, we remove key steps and inject irrelevant content to break the logical flow; for T4-type chains, we apply both T2-type and T3-type mutations. Each variant is created for 50 tasks to ensure balanced coverage across the four types.

\begin{figure}[t]
    \centering
    \includegraphics[width=0.49\textwidth]{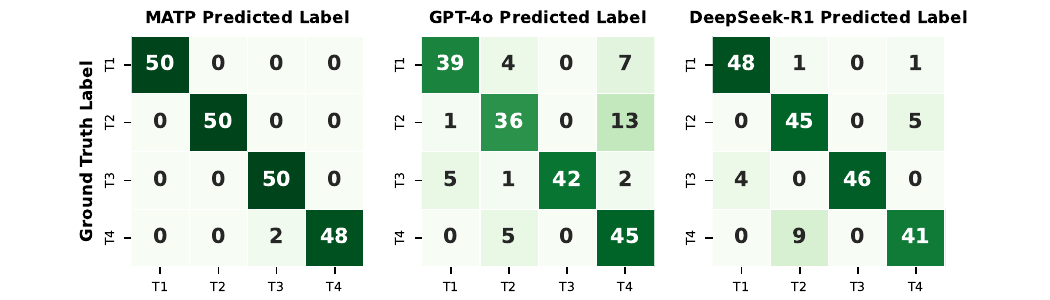}\vspace{-0.3cm}
    \caption{Heatmap of Reasoning Chain Classification Performance for \toolname{} and Baselines.}\vspace{-0.3cm}
    \label{fig:rq2}
\end{figure}

\noindent \textbf{Results.}
\autoref{fig:rq2} presents the classification accuracy of \toolname{} versus the baselines. As shown, \toolname{} achieves 100\% accuracy on T1–T3 and 96\% on T4. A review of the two T4 misclassifications attributes the errors to minor NL2FOL translation deviations, suggesting that with correct translation, \toolname{} can consistently classify reasoning chains accurately. 
In contrast, the GPT-4o-based baseline achieves lower accuracy—78\%, 72\%, 84\%, and 90\% for T1–T4, respectively. Error analysis reveals two main issues: (1) 54 correct steps were misjudged as incorrect, causing 14 T1/T3 cases to be classified as T2/T4; and (2) it struggles to assess whether steps sufficiently support the conclusion, leading to 20 T1/T2 cases misclassified as T3/T4 and 11 T3/T4 cases misclassified as T1/T2. 
The DeepSeek-R1-based baseline performs better, achieving 96\%, 90\%, 92\%, and 82\% accuracy. Most misclassifications arise from failure to assess whether steps sufficiently support conclusions—6 T1/T2 cases were mistaken for T3/T4, and 13 T3/T4 as T1/T2—mainly due to its weaker instruction-following ability, often defaulting to global premise-to-conclusion matching.

While 100\% classification accuracy can be guaranteed with perfect translation, we cannot ensure flawless NL2FOL translation in practice. 
Consequently, \toolname{} cannot ensure absolute correctness in real-world applications. Nevertheless, it achieves significantly higher accuracy than baselines.
We also record the average processing time per task for all tools. GPT-4o takes around 6s per task, DeepSeek-R1 is significantly slower at 96s, while \toolname{} maintains a comparable speed of approximately 9s.
\vspace{-1mm}
\begin{tcolorbox}[title=ANSWER to RQ2, boxrule=0.8pt,boxsep=1.5pt,left=2pt,right=2pt,top=2pt,bottom=1pt,fontupper=\small]
\toolname{} achieves perfect classification given accurate NL2FOL translation, clearly outperforming the prompting-based baselines.
\end{tcolorbox} 
\vspace{-2mm}

\subsection{RQ3:~Model Reasoning Analysis}


Using the reasoning chain classifications produced by \toolname{}, we evaluate how different LLMs perform across tasks, focusing on reasoning quality and model-specific patterns. 
We first compare model performance across datasets, identifying strengths, limitations, and reasoning tendencies.
Next, we analyze structural patterns within each category to uncover common behaviors and failure modes. 
Lastly, we examine cases where \toolname{}  fails to evaluate and summarize typical causes. 
This analysis reveals key reasoning bottlenecks and guides future improvements in LLM reasoning.


\begin{figure*}[t]
    \centering
    \includegraphics[width=0.9\linewidth]{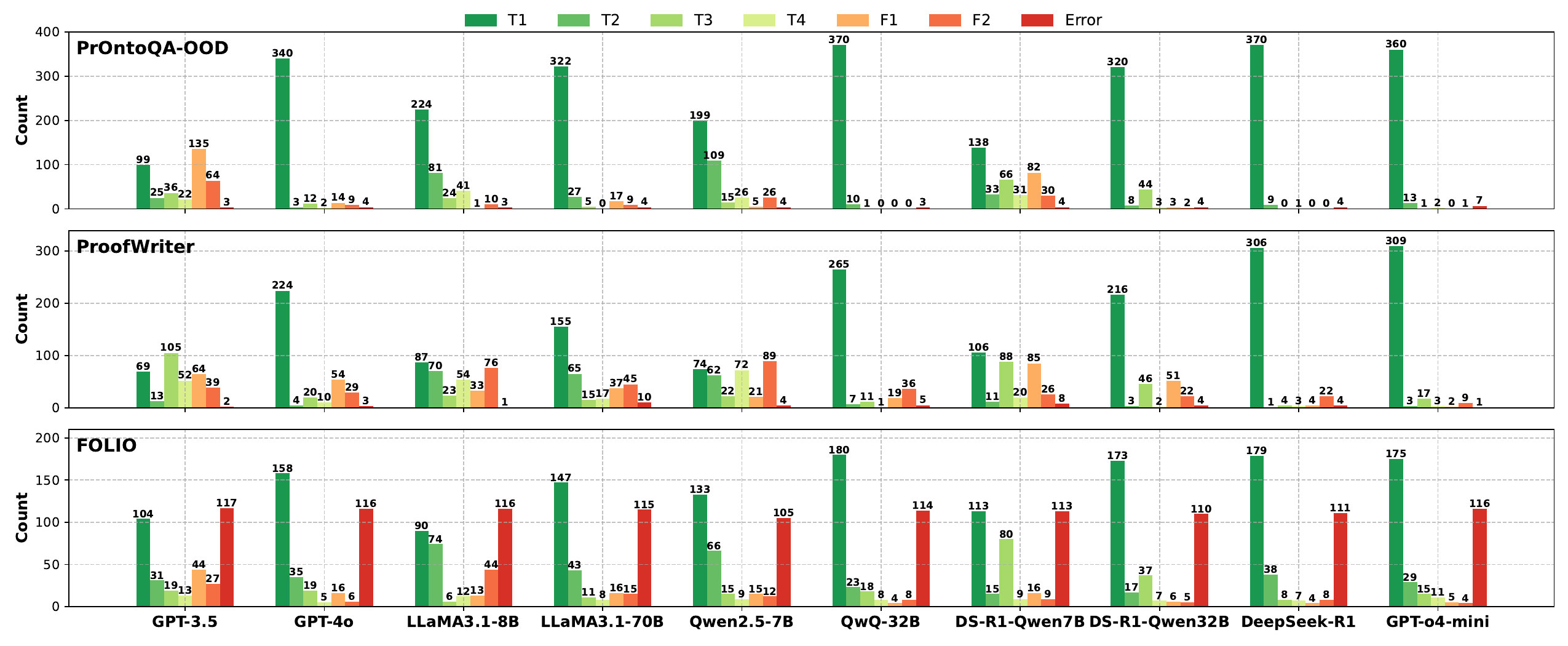}\vspace{-0.4cm}
    \caption{Category Distribution of LLM-Generated Reasoning Chains on Benchmark Datasets.}\vspace{-0.3cm}
    \label{fig:rq3_2}
\end{figure*}

\subsubsection{Analysis of Model Performance} 
The overall category distribution is shown in \autoref{fig:rq3_2}. The Error part indicates cases where \toolname{} fails to evaluate. 
The failure rate of MATP analysis cases averages 1.04\% on \prontoqa{}, 1.22\% on \proofwriter{}, and 31.92\% on \folio{}.
The notably higher rate on \folio{} suggests that \toolname{} is more sensitive to complex, natural language structures, mainly due to challenges in NL2FOL translation.
The specific failure reasons are discussed in \autoref{sec:rq3_3}.
These failed cases are excluded from subsequent classification analysis. 

While reasoning and general models achieve similar answer accuracy on simple datasets like \prontoqa{}, they differ significantly in actual reasoning quality.
Reasoning models (e.g., DeepSeek-R1, QwQ-32B) consistently generate valid reasoning chains (T1–T2), with DeepSeek-R1 reaching over 96\% T1 on \textsc{PrOnt\-oQA-OOD}. 
In contrast, general models (e.g., LLaMA3.1–8B) often rely on shallow or inconsistent logic, lacking valid proof paths despite high accuracy.
General models also show issues like redundant logic (LLaMA) or overuse of irrelevant premises (GPT-3.5), leading to correct but weakly justified answers. 
Even reasoning models struggle on challenging datasets like \proofwriter{}, where incomplete or invalid chains (T3/T4) still exist.
On \folio{}, the proportion of T2 and T4 increases across all models. This stems partly from the evaluation tool's difficulty in handling truly natural language, and partly from the dataset’s reliance on background knowledge not always captured in the explicit premises.
Model scale also affects performance. Larger models such as LLaMA3.1–70B generally yield higher T1 ratios and fewer errors than smaller variants, yet still lag behind reasoning-specialized models in logical coherence.

In summary, answer accuracy alone fails to capture reasoning strength. Our classification strategy reveals critical differences in logical validity and robustness, highlighting the need to evaluate reasoning processes, not just final answers.

\begin{table}[!htbp]
\centering
\caption{Special Cases of Reasoning Chain Categorization with Representative Examples. In the Reasoning Response, sentences highlighted in \textcolor{orange}{orange} are considered $Unknown$ relative to the task premises, while black sentences are $True$.}\vspace{-0.3cm}
\label{table:rq3_2}
\renewcommand{\arraystretch}{1.2} 
\fontsize{9pt}{12pt}\selectfont
\resizebox{0.48\textwidth}{!}{
\begin{tabular}{p{4cm}|p{6.5cm}|p{2.5cm}} 
\hline
\multicolumn{1}{c|}{\textbf{Description}} & \multicolumn{1}{c|}{\textbf{Reasoning Response}} & \multicolumn{1}{c}{\textbf{Task Name}} \\
\hline
\textbf{T2:} If the $Unknown$ step is excluded, this answer is a standard response. &  
\textbf{GPT-4o:} 1) The tiger eats the dog. 
2) If someone eats the dog then the dog needs the bear. 
3) \textcolor{orange}{The dog eats the dog.}
4) The dog needs the bear. 
5) There is evidence to conclude that the dog needs the bear. & 
\textbf{Task \scalebox{1.1}{\ding{172}}:} \newline \proofwriter{}-RelNeg-OWA-D5-226-Q3 
\\
\hline
\textbf{T3:} The conclusion that needs to be proven is ``Fiona is not big.''. However, the model is proving ``Anne is big.''. & 
\textbf{GPT-3.5:} ... 5) If Anne is white then Anne is big.
6) There is evidence to conclude that Anne is big. &
\textbf{Task \scalebox{1.1}{\ding{173}}:} \newline
\proofwriter{}-AttNeg-OWA-D5-1003-Q5 
\\
\hline
\textbf{T4:} The model introduce an $Unknown$ information to fabricate a valid proof path. & 
\textbf{DS-R1-Qwen7B:}
1) Sally is a dumpus.
2) Sally is a gorpus.
3) Every gorpus is a zumpus.
4) Therefore, Sally is a zumpus.
5) \textcolor{orange}{Every zumpus is a romrus.}
6) Therefore, Sally is a romrus. &
\textbf{Task \scalebox{1.1}{\ding{174}}:} \newline
\prontoqa{}-4hop-Composed-random-noadj-69-4
\\
\hline
\textbf{F1:} The model's reasoning process contains sufficient information to derive the conclusion, yet it incorrectly responds that it cannot. &
\textbf{QwQ-32B:}
1) Fiona is quiet.
2) Quiet things are green.
3) Fiona must be green.
4) Fiona is rough.
... 
9) Therefore, Fiona is not green. &
\textbf{Task \scalebox{1.1}{\ding{175}}:} \newline
\proofwriter{}-AttNeg-OWA-D5-630-Q4 
\\
\hline
\textbf{F1:} The model's reasoning steps are entirely correct, but the final answer is incorrect. & 
\textbf{GPT-4o:}
1) Bob is quiet.
2) Bob is smart.
3) All smart things are green.
4) Bob is green.
5) Green things are white.
6) Bob is white.
7) There is evidence to conclude that Bob is white. &
\textbf{Task \scalebox{1.1}{\ding{176}}:} \newline \proofwriter{}-AttNeg-OWA-D5-327-Q5
\\
\hline
\textbf{F1:} The model enters a loop during reasoning, causing the inference process to stall. &
\textbf{LLaMA3.1-8B:}
1) Every lorpus is a numpus.
2) Every lorpus is a numpus.
...
20) Every lorpus is a numpus.
... &
\textbf{Task \scalebox{1.1}{\ding{177}}:} \newline \prontoqa{}-4hop-Composed-random-noadj-22-7 
\\
\hline

\end{tabular}}\vspace{-3mm}
\end{table}

\subsubsection{Analysis of Reasoning Chain Categories}
We further analyze each category to identify common patterns and provide representative examples in \autoref{table:rq3_2}.

\noindent \textbf{T1} chains that are fully correct in both step-level logic and overall consistency, representing ideal outputs. Nonetheless, even when satisfying our classification criteria, there remain minor issues that warrant further discussion,
such as redundant steps that are correct but irrelevant, and skipped steps where a step relies on multiple unstated premises. Such cases are illustrated in GPT-4o’s output in \autoref{fig:motivation} (step 1 and step 4).

\noindent \textbf{T2} responses resemble T1 but contain one or more incorrect steps that do not affect the final conclusion. Once removed, the remaining steps form an entirely correct and valid proof (i.e., task \scalebox{1.1}{\ding{172}}).

\noindent \textbf{T3} chains consist of correct steps but fail to form a valid proof path. This often results from implicit reasoning, where key premises are omitted, or accidental correctness, where an unrelated conclusion is proven that happens to share the same truth value (i.e., task \scalebox{1.1}{\ding{173}}). Models often default to answering $False$ when uncertain, which may also produce coincidentally correct answers.

\noindent \textbf{T4} exhibits the same issues as T3 but also includes incorrect reasoning steps. These outputs are typically disorganized and incoherent. Some cases even introduce non-existent facts to fabricate plausible chains (i.e., task \scalebox{1.1}{\ding{174}}).

\noindent \textbf{F1} and \textbf{F2} responses both yield incorrect final predicted answers. Reasoning chains in F1 contain only correct steps, whereas in F2 include at least one incorrect step. 
Common issues include incomplete reasoning due to missing or poorly integrated premises (i.e., task \scalebox{1.1}{\ding{175}}). We also observe special failure modes: models may derive the correct conclusion but answer it incorrectly (i.e., task \scalebox{1.1}{\ding{176}}), or enter repetitive loops that stall the reasoning process (i.e., task \scalebox{1.1}{\ding{177}}).

\begin{table}[!htbp]
\centering
\caption{\toolname{} Misclassifications on DeepSeek-R1's Response for folio-train-57-168. \textcolor{red}{Red} indicates \toolname{}’s erroneous outputs, while \textcolor{newgreen}{green} marks the expected correct ones.}\vspace{-0.3cm}
\label{table:rq3_2_mislassifications}
\renewcommand{\arraystretch}{1.2} 
\fontsize{9pt}{12pt}\selectfont
\resizebox{0.48\textwidth}{!}{
\begin{tabular}{p{5.5cm}|p{5cm}|p{2cm}} 
\hline
\multicolumn{1}{c|}{\textbf{Task Info and Reasoning Response}} & \multicolumn{1}{c|}{\textbf{Step-FOL Generated by GPT-4o}} & \multicolumn{1}{c}{\textbf{\toolname{} Output}} \\
\hline
\textbf{Conclusion: } Leo is an animal. (True) \newline
\textbf{Filtered Resoning Steps:} \newline1) All pets are animals. \newline2) Charlie has a naughty pet dog named Leo. \newline3) A pet dog is a type of pet. \newline4) Since all pets are animals, Leo being a pet implies Leo is an animal. 
&  
1) \begin{math}
    \forall x (pet(x) \rightarrow animal(x))
\end{math} \newline
2) \begin{math}
    pet(leo) \land dog(leo) \land naughty(leo) \land haspet(charlie, leo)
\end{math} \newline
\textcolor{red}{3)} \begin{math}
    \textcolor{red}{\forall x (dog(x) \rightarrow pet(x))}
\end{math} \newline
\textcolor{newgreen}{3)} \begin{math}
    \textcolor{newgreen}{\forall x (pet(x) \land dog(x) \rightarrow pet(x))}
\end{math} \newline
4) \begin{math}
    \forall x ((pet(x) \land animal(x)) \rightarrow (pet(leo) \rightarrow animal(leo)))
\end{math}
& 
S1: True
\newline
S2: True
\newline
\textcolor{red}{S3: Unknown}
\newline
\textcolor{newgreen}{S3: True}
\newline
S4: True
\newline
\textcolor{red}{Cate: T2}
\newline
\textcolor{newgreen}{Cate: T1}
\\

\hline

\end{tabular}}\vspace{-1.5mm}
\end{table}

In summary, T1 and T2 exhibit the most reliable reasoning, while T3 and T4 are less stable. F1 shows potential for improvement, while F2 is the most flawed. Overall reasoning quality follows: \textbf{T1 $>$ T2 $>$ T3 $>$ T4 $>$ F1 $>$ F2}.
Manual inspection also reveals a few misclassifications. As shown in \autoref{table:rq3_2_mislassifications}, an NL2FOL translation error in one step leads to T1 being mislabeled as T2.
While the classification is not perfectly accurate, it reliably captures overall reasoning patterns and pinpoints potential error steps, offering clearer consistency and interpretability than prompt-based baselines.

\begin{table}[!htbp]
\centering
\caption{Representative Error Cases in NL2FOL Translation. Key erroneous parts are highlighted in \textcolor{red}{red}.}
\vspace{-0.3cm}
\label{table:rq3_3}
\renewcommand{\arraystretch}{1.2} 
\fontsize{9pt}{11pt}\selectfont
\resizebox{0.48\textwidth}{!}{
\begin{tabular}{p{3.7cm}|p{3.2cm}|p{4.1cm}|p{1.8cm}} 
\hline
\multicolumn{1}{c|}{\textbf{Description}} & \multicolumn{1}{c|}{\textbf{NL Sentences}} & \multicolumn{1}{c|}{\textbf{FOL Representation}} & \multicolumn{1}{c}{\textbf{Dataset}} \\
\hline
The domain of the universal quantifier in FOL is wrong. & If someone is smart and not kind then they are young.  &
\begin{math}
    \forall x \textcolor{red}{(} smart(x)\land\neg kind(x)\textcolor{red}{)}\rightarrow young(x)
\end{math} & 
Case \scalebox{1.2}{\ding{182}} from \proofwriter{}
\\
\hline
The wrong use of $\land$ instead of $\lor$ for ``or''. & 
Alex is a gorpus, a sterpus, a numpus, \textcolor{red}{or} a gorpus.  & 
\begin{math}
    Gorpus(alex) \textcolor{red}{\land}  Sterpus(alex)  \linebreak \textcolor{red}{\land} Numpus(alex) \textcolor{red}{\land} Gorpus(alex)
\end{math} & 
Case \scalebox{1.2}{\ding{183}} from \prontoqa{}
\\
\hline
$BoyBand(btob4u)$ which is effective for reasoning candidate conclusion is omitted. & 
Show Your Love is a song recorded by the South Korean \textcolor{red}{boy band} BtoB 4u.  & 
\begin{math}
    Song(showyourlove) \land RecordedBy(showyourlove, \linebreak btob4u)
\end{math} & 
Case \scalebox{1.2}{\ding{184}} from \folio{}
\\
\hline
The predicate is used improperly, where the same word ``referee observer'' in the context is represented by different predicates. 
& Brian Winter was appointed as a \textcolor{red}{referee observer} after his retirement. Some football referees become \textcolor{red}{referee observer}s.
& $\textcolor{red}{AppointedObserver}(brian).$ $\exists x (FootballReferee(x) \land \textcolor{red}{RefereeObserver}(x))$
& Case \scalebox{1.2}{\ding{185}} from \folio{}
\\
\hline
``Either...or...'' represents XOR, which differs from OR and should be denoted using the $\oplus$.
& Everything displayed in the collection is \textcolor{red}{either} a plant \textcolor{red}{or} an animal. 
& $\forall x (Displayed(x) \rightarrow (Plant(x) \textcolor{red}{\lor} Animal(x)))$
& Case \scalebox{1.2}{\ding{186}} from \folio{}
\\
\hline
 The direct use of $\textgreater$ is not grammatical and should be expressed as $MoreRevenue(airbus, boeing)$.
& Airbus made \textcolor{red}{more} revenue \textcolor{red}{than} Boeing last year.
& \begin{math}
    Revenue(airbus, lastyear) \textcolor{red}{\textgreater} \linebreak
    Revenue(boeing, lastyear)
\end{math}
& Case \scalebox{1.2}{\ding{187}} from \folio{}


\\
\hline
\end{tabular}}\vspace{-0.3cm}
\end{table}

\subsubsection{Analysis of Error Cases}
\label{sec:rq3_3}
Most evaluation failures originate from NL2FOL translation errors. \autoref{table:rq3_3} presents representative cases. While translation from \prontoqa{} and \proofwriter{} is largely accurate, occasional issues remain, such as incorrect quantifier scope (i.e., case \scalebox{1.1}{\ding{182}}) or misinterpreting ``or'' as a conjunction when linking multiple entities (i.e., case \scalebox{1.1}{\ding{183}}). In contrast, \folio{} introduces more diverse and complex errors due to its rich natural language. We summarize common error patterns as follows.


\noindent \textbf{Information Omission.} 
Key details are omitted when sentences contain multiple facts (case \scalebox{1.1}{\ding{184}}). 

\noindent \textbf{Improper Predicate Usage.}
The same concept may be mapped to different predicates, causing semantic inconsistency (i.e., case \scalebox{1.1}{\ding{185}}).

\noindent \textbf{XOR vs. OR Confusion. } 
Expressions involving exclusive choice, such as ``either...or...'', are often mistakenly represented as disjunction ($\lor$) rather than exclusive or ($\oplus$), even when explicitly prompted (i.e., \scalebox{1.1}{case \ding{186}}).

\noindent \textbf{Inadequacy in Handling Domain-Specific Expressions.} 
Domain-specific expressions involving comparatives, superlatives, mathematics, or time are often poorly handled, resulting in ungrammatical FOL (i.e., case \scalebox{1.1}{\ding{187}}).

\begin{figure}[ht]
    \centering    \includegraphics[width=0.48\textwidth]{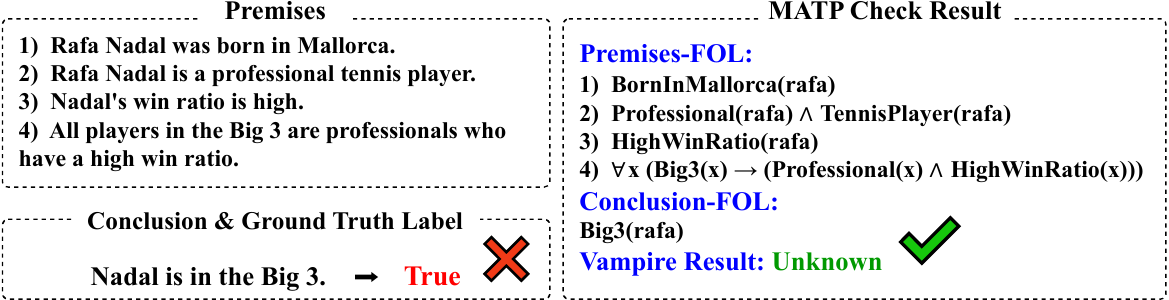}\vspace{-0.3cm}
    \caption{Example of Annotation Error in \folio{}. The task ID is folio-train-34-99.}\vspace{-0.3cm}
    \label{fig:annotation_error}
\end{figure}

\noindent \textbf{Annotation Errors.} In addition to the extraction issues above, we also observe annotation errors in the \folio{} dataset during manual inspection. \autoref{fig:annotation_error} provides an example of such a mislabeling. We find that the FOL ground truth representation of the fourth premise in the task is incorrectly given as ``\begin{math}
    \forall x ((Professional(x) \land HighWinRatio(x)) \rightarrow Big3(x))
\end{math}'', which may be the cause of the labeling error.
Similar issues are also found in tasks such as folio-train-63-186, 167-480, 167-479, 302-746, 470-1359, and 470-1360.

\vspace{-1mm}
\begin{tcolorbox}[title=ANSWER to RQ3, boxrule=0.8pt,boxsep=1.5pt,left=2pt,right=2pt,top=2pt,bottom=1pt,fontupper=\small]
\toolname{} reveals clear differences in reasoning behavior across LLMs. Despite similar answer accuracy, reasoning models yield more coherent logic than general models, highlighting the need to assess reasoning quality beyond final answers.
\end{tcolorbox} 
\vspace{-2mm}

\subsection{Ablation Study}
\label{sec:AblationStudy}
We conduct an ablation study comparing \toolname{}’s performance with and without feedback during NL2FOL regeneration.

\noindent \textbf{Setup.} 
As shown in \autoref{fig:rq3_2}, most analysis failures occur on \folio{}. Therefore, we evaluate the responses of two models, GPT-4o and DeepSeek-R1, on 355 tasks from \folio{}. 
Since pinpointing specific errors is difficult, we use the entire previous generation as feedback.
We extend the original prompt in \autoref{fig:nlfol_prompt} by adding a \textit{Feedback} field under \textit{Input instance}, containing the prior NL2FOL translation with conversion errors.
This field helps the model avoid similar mistakes during regeneration. Similarly, the maximum number of regeneration attempts is set to three. 
We focus on the number of failures in both feedback and non-feedback settings.

\noindent \textbf{Results.} As shown in \autoref{table:ablation_study}, using Feedback has minimal effect, with non-feedback settings sometimes yielding better results. Additionally, feedback increases resource consumption. 
Therefore, feedback is excluded during MATP’s actual implementation.

\vspace{-2mm}
\begin{table}[!htbp]
\centering
\fontsize{9}{11}\selectfont
\caption{Impact of Feedback on \toolname{} Failures.}
\vspace{-0.3cm}
\label{table:ablation_study}
\begin{tabular}{ccc}
\hline
\textbf{Model}     & \textbf{With Feedback}   &\textbf{Without Feedback} \\
\hline
\textbf{GPT-4o } & 118 & 116 \\
\textbf{DeepSeek-R1} & 119 & 111 \\
\hline
\end{tabular}
\end{table}
\vspace{-2mm}

\section{Discussion}\label{sec:discussion}
\subsection{Future Enhancement}

This paper proposes (1) a framework for translating informal LLM reasoning into formal logic and verifying with ATP tools, and (2) a multi-level classification strategy beyond binary answer correctness. 
Thanks to the modular design of \toolname{}, we can independently upgrade each component, making it adaptable for deployment in high-stakes scenarios that require rigorous reasoning validation.
Several directions remain for future exploration:

\noindent\textbf{Enhanced Formalization.} \toolname{}'s accuracy depends on reliable NL2FOL translation. 
Current prompt-based methods still show limitations with complex natural language inputs. 
Future work may incorporate more robust or domain-adapted translation tools.

\noindent \textbf{Open-Domain Knowledge Requirement.}
Some mappings (e.g., ``Low'' to ``$\neg$ High'') require external knowledge not present in the premises.
To ensure a strict one-to-one mapping between natural and formal language while handling implicit common knowledge, the model needs to autonomously introduce necessary knowledge (e.g., ``low'' equals ``not high'') either before or during translation.

\noindent\textbf{Improved Evaluation Metrics.} Evaluation of extracted FOL typically relies on gold-standard annotations and similarity-based metrics, which often fail to reflect semantic fidelity and logical validity~\cite{thatikonda2025assessing}. This underscores the need for more principled and automated evaluation methods.

\noindent\textbf{Reconstructing Reasoning Graphs.} Future work may model reasoning steps as nodes and ATP-detected inferences as directed edges, constructing a reasoning graph from LLM-generated reasoning outputs. This structure enables fine-grained verification and helps identify redundant, missing, or conflicting logic beyond simple path existence checks.


\noindent\textbf{Expanded Reasoning Support.} Beyond forward entailment, supporting reverse reasoning could broaden applicability.




\subsection{Optimization on Response Analysis}
To improve efficiency and reduce manual effort—particularly in NL2FOL evaluation, we leverage LLM internal features using entropy and Jensen-Shannon (JS) divergence~\cite{MENENDEZ1997307} to identify uncertain or potentially flawed reasoning steps. As demonstrated in \autoref{fig:entropy}, when Llama3.1-8B series are evaluated on \proofwriter, flawed reasoning often correlates with higher entropy or JS divergence. Note that the full figure across datasets and LLMs is available in ~\cite{matp}.

Based on this, we implement a threshold-based filtering strategy: reasoning outputs with entropy below 7.0 and JS divergence below 0.065 are considered likely correct, allowing us to focus analysis on more uncertain cases.
This pre-filtering step effectively reduces the evaluation burden, with correct reasoning coverage rates ranging from 14.7\% to 40.5\% across datasets and models.
While some incorrect responses are mistakenly filtered out (e.g., 8.5\% for \proofwriter and 6.7\% for \folio under Llama3.1-8B), other configurations show near-zero false negatives, validating the approach’s reliability.
Overall, this method improves evaluation efficiency by prioritizing uncertain reasoning steps.
Future work will explore tighter integration of LLM internal signals to enable automated filtering and semi-supervised reasoning quality assessment.

\begin{figure}[h]
    \centering\vspace{-0.1cm}
    \includegraphics[width=0.9\linewidth]{fig/entropy_vs_js_divergence.pdf}\vspace{-0.3cm}
    \caption{Comparison of entropy and Jensen–Shannon divergence distributions for correct (T) and incorrect (F) reasoning responses for Llama3.1 series on \proofwriter. }
    \vspace{-0.5cm}
    \label{fig:entropy}
\end{figure}

\section{Related Work}

\noindent\textbf{Automatically Generating Formal Specifications.}
LLMs have been increasingly applied to automate formal specification generation and logical translation from natural language. 
SpecGen~\cite{ma2024specgen} adopts a conversational and mutation-based approach to generate program specifications without relying on templates. NL2CTL~\cite{10.1007/978-981-96-0617-7_1} translates requirements into Computation Tree Logic~(CTL) by constructing an NL-CTL dataset through prompt engineering and fine-tuning a T5 model.
SAFE~\cite{chen2024automated} targets Rust, integrating data synthesis, fine-tuning, and symbolic verification to automate proof generation with over 70\% accuracy. NL2FOL~\cite{Lalwani2024NL2FOLTN} converts natural language into FOL, using SMT solvers for fallacy detection and explanation. NL2SPEC~\cite{cosler2023nl2spec} supports interactive translation into temporal logic, allowing iterative refinement to resolve ambiguity.

\noindent\textbf{Enhancing Reasoning Capabilities of LLMs.}
Enhancing LLM reasoning is critical for high-stakes tasks.
Internally, DeepSeek-R1~\cite{guo2025deepseek} combines supervised fine-tuning and reinforcement learning to enable reasoning abilities such as chain-of-thought and self-verification.
Externally, prompting strategies like Logic-of-Thought~\cite{liu2024logic} and Chain-of-Knowledge~\cite{wang-etal-2024-boosting-language} improve logical completeness and factual coherence by injecting structured logic or knowledge into prompts.
Tool-augmented methods improve reasoning accuracy by using LLMs as semantic parsers to translating natural language into formal languages, which are subsequently processed by external symbolic tools.
LINC~\cite{olausson-etal-2023-linc} converts NL into FOL for logic theorem proving.
SatLM~\cite{NEURIPS2023_8e9c7d4a} translates NL into declarative task specifications for Z3 SMT solver~\cite{de2008z3}, targeting mathematical and logical reasoning tasks.
Logic-LM~\cite{pan2023logiclm} covers four categories of formal languages and corresponding tools, selecting appropriate ones for different types of reasoning tasks.
Our work adopts a similar strategy but focuses on evaluating the correctness and logical consistency of LLM responses to reasoning tasks.


\noindent 
\textbf{Evaluating Reasoning Capabilities of LLMs.}
Evaluating LLM reasoning has become a key research focus.
Prior studies assess model performance across diverse tasks, such as code understanding correctness at both the function~\cite{li2024mutation} and software engineering~\cite{tanzil2024chatgpt} levels, as well as factual hallucination detection in real-world reasoning scenarios~\cite{li2024drowzee}.
As reasoning abilities advance, evaluation objectives have expanded beyond overall task correctness to step-by-step reasoning analysis.
The survey~\cite{lee2025evaluating} summarizes four key dimensions for evaluating reasoning steps: \textit{factuality}, \textit{validity}, \textit{coherence}, and \textit{utility}.
Despite progress, most methods either rely on labor-intensive gold-standard reasoning chains~~\cite{golovneva2022roscoe, PrOntoQA} or probabilistic LLM-based evaluations~\cite{zeng2024mrgsm8kmetareasoningbenchmarklarge, xia2024evaluatingmathematicalreasoning}, which undermine logical rigor.
ROSCOE~\cite{golovneva2022roscoe} proposes several metrics using embedding-based calculations among reasoning steps, reference steps, and problems for evaluation.
MR-GSM8K~\cite{zeng2024mrgsm8kmetareasoningbenchmarklarge} prompts LLMs to check reasoning correctness, outputting structured fields like \textit{Solution Correctness}, \textit{First Error Step}, and \textit{Error Analysis}. 
ReasonEval~\cite{xia2024evaluatingmathematicalreasoning} fine-tunes LLMs on high-quality labeled data to score \textit{validity} and \textit{redundancy} for each reasoning step.
Our work extends this research line by incorporating automated theorem provers to evaluate LLM-generated reasoning steps.

\vspace{-1mm}
\section{Conclusion}

In this paper, we presented MATP, an evaluation framework that bridges the gap between informal LLM reasoning and formal verification through automated theorem proving. By systematically converting natural language reasoning into First-Order Logic and applying rigorous verification at each step, MATP provides unprecedented insight into the logical validity of LLM-generated reasoning chains. Our extensive evaluation demonstrates that MATP not only achieves high accuracy in FOL translation and reasoning verification, but also enables fine-grained classification that distinguishes genuine deductive reasoning from superficial pattern matching or accidental correctness. The stark differences revealed between models underscore the critical need for formal verification in high-stakes applications. While challenges remain in handling complex linguistic constructs, MATP establishes a foundational framework for ensuring reliable, verifiable AI reasoning in safety-critical domains where trust through logical soundness is paramount.


\bibliographystyle{ACM-Reference-Format}
\bibliography{9.ref}



\end{sloppypar}
\end{document}